# Treatment of Unicode canoncal decomposition among operating systems

Efstratios Rappos
Efstratios.Rappos@heig-vd.ch
University of Applied Sciences and Engineering of Western Switzerland
HEIG-VD, Route de Cheseaux 1, 1400 Yverdon-les-Bains, Switzerland

24 April 2015

**Abstract**

This article shows how the text characters that have multiple representations under the Unicode standard are treated by popular operating systems. Whilst most characters have a unique representation in Unicode, some characters such as the accented European letters, can have multiple representations due to a feature of Unicode called normalization. These characters are treated differently by popular operating systems, leading to additional challenges during interoperability of computer programs.

## 1  Introduction

Unicode [1] specifies that some characters can be represented by more than one Unicode sequences. For example, the character 'é' can be represented as a precomposed or decomposed character:

| Unicode | é | e | ´ |
|---|---|---|---|
|  | LATIN SMALL LETTER E WITH ACUTE | LATIN SMALL LETTER E | COMBINING ACUTE ACCENT |
| sequence | U+00E9 | U+0065 | U+0301 |
| output | é | é | |

Table 1: An example of a character with two Unicode representations

In Table 1 the first representation is a precomposed character whereas the second is a decomposed character consisting of two Unicode codepoints. However, some characters can have more than two choices for representation, such as the character Å, as shown in Table 2. All these characters are canonically equivalent under Unicode.

| Å | Å | A | ° |
|---|---|---|---|
| LATIN CAPITAL LETTER A WITH RING ABOVE | ANGSTROM SIGN | LATIN CAPITAL LETTER A | COMBINING RING ABOVE |
| U+00C5 | U+212B | U+0041 | U+030A |

Table 2: An example of a character with multiple Unicode representations

The same is true for in the case of multiple accents as in ắ, shown in Table 3:

| ǻ | å | ´ | a | ˚ | ´ |
|---|---|---|---|---|---|
| LATIN SMALL LETTER A WITH RING ABOVE AND ACUTE | LATIN SMALL LETTER A WITH RING ABOVE | COMBINING ACUTE ACCENT | LATIN SMALL LETTER A | COMBINING RING ABOVE | COMBINING ACUTE ACCENT |
| U+01FB | U+00E5 | U+0301 | U+0061 | U+030A | U+0301 |

Table 3: An example where two codepoints represent equivalent characters

These complexities and ambiguities get worse in the case of Icelandic, Greek or Vietnamese scripts, with some examples shown in Table 4 and Table 5. Finally, there can be cases where a precomposed character does not exist in Unicode, so the character can only be represented as a combination of a base character and accents as combining characters.

| ǭ | ǫ | ̄ | o | ̄ | ̨ | o | ̨ | ̄ | ō | ̨ |
|---|---|---|---|---|---|---|---|---|---|---|
| LATIN SMALL LETTER O WITH OGONEK AND MACRON | LATIN SMALL LETTER O WITH OGONEK | COMBINING MACRON | LATIN SMALL LETTER O | COMBINING MACRON | COMBINING OGONEK | LATIN SMALL LETTER O | COMBINING OGONEK | COMBINING MACRON | LATIN SMALL LETTER O WITH MACRON | COMBINING OGONEK |
| U+01ED | U+01EB | U+0304 | U+006F | U+0304 | U+0328 | U+006F | U+0328 | U+0304 | U+014D | U+0328 |

Table 4: An example of a character with multiple choices for the order of the accents

| ᾤ | ᾤ | ͅ | ὤ | ´ | ͅ | ω | ᾽ | ´ | ͅ | ... |
|---|---|---|---|---|---|---|---|---|---|---|
| GREEK SMALL LETTER OMEGA WITH PSILI AND OXIA AND YPOGEGRAMMENI | GREEK SMALL LETTER OMEGA WITH PSILI AND OXIA | COMBINING GREEK YPOGEGRAMMENI | GREEK SMALL LETTER OMEGA WITH PSILI | COMBINING ACUTE ACCENT | COMBINING GREEK YPOGEGRAMMENI | GREEK SMALL LETTER OMEGA | COMBINING COMMA ABOVE | COMBINING ACUTE ACCENT | COMBINING GREEK YPOGEGRAMMENI | ... |
| U+1FA4 | U+1F64 | U+0345 | U+1F60 | U+0301 | U+0345 | U+03C9 | U+0313 | U+0301 | U+0345 | ... |

Table 5: An example of a character with multiple choices for the order of the accents (2)

The next sections show the choice that each operating system and filesystem performs when deciding which representation of an accented character to use.

## 2 Treatment in the NTFS filesystem

The NTFS filesystem accepts all valid Unicode characters in filenames except for the characters '\', '/', ':', '?', '*'. The period is allowed as an extension separator in the middle of the filename only. Under Windows, further restrictions include '<', '>', ':', '|', '"', ASCII characters 0-31 as well as length restrictions and special sequences such as 'com1', 'prn' and 'nul' [3].

In order to determine the normalization procedure during file creation, we distinguish between the following procedures:

- creating a file via the Windows Explorer interface
- creating a file via a batch file in the command line
- creating a file programmatically in Java or Python

**Creating a file via the Windows Explorer interface**    When creating a file via the visual Windows Explorer interface and giving a name by typing in accented characters from the keyboard, the operating system stores the character as a precomposed character.

**Creating a file via the command line**   As with the , when creating a file via the command line or a batch file, the system uses the precomposed form of those characters in the filename.

**Creating a file programmatically in Java and Python**   In this case, the operating system stores whatever sequence of unicode characters is used as the input. No conversion is made by the operating system (Windows) or the filesystem (NTFS). This means that both files can co-exist and the existence of one does not prevent the other from being created.

## 3   Treatment in the Mac OS HFS+ filesystem

In Macintosh HFS+ filesystem the only illegal character is ':'. Other limitations exist, such as the use of '/' as a separator in the BSD-style terminal [2]. However, unlike the other operating systems, the filenames are stored as a decomposed unicode (a slightly modified version of Unicode NFD decomposition). So a filename that contains the character 'é' will be stored as two Unicode points using the NFD decomposition. If the filename exists with one Unicode representation and an attempt is made to create the same file with the alternative decomposition, the first file will be overwritten or an error will occur 'file already exists'. The NFD decomposition is done at a very low level of the operating system and cannot be bypassed.

We note also that the Mac OS X file system uses a modified NFD decomposition when converting precomposed Unicode to the canonical decomposition. In that scheme, some character ranges remain composed in spite of the NFD rules. An example is the Angstrom sign, if a filename contains this character it will remain as such and not be decomposed as the Unicode NFD rules would require.

## 4   Treatment in the Linux Ext4 filesystem

In the Linux operating system under the ext4 filesystem no conversion takes place. The filename is saved in the same manner as it is input. In this case, two different files can co-exist, one with a particular decomposition and one with the other one.

## 5   Interoperability

The following issues relating to interoperability between filesystems have been identified:

- If a filename 'nul' or 'file/' is created on Mac and moved to Windows it will not be viewable or editable.

- If a file containing precomposed characters is created in Windows or Linux and then transferred to Mac, it will be converted to decomposed NFD Unicode.

- If an attempt is made in Mac OS X to access a file using the precomposed method, it will return the decomposed file if it exists.

Of course, a lot of this depends on the method used to transfer the file (e.g., via a external FAT32 Usb stick, via a Samba connection or via an FTP or SCP connection) as some conversion of illegal input is done by these tools and not by the underlying Operating System. However, regardless of the method used, Mac OS X does not allow the simultaneous presence of two files which have filenames consisting of different Unicode sequences but are canonically equivalent.

# 6 Conclusions

This short article presents some important differences in the treatment of Unicode characters that have multiple, canonically-equivalent representations among the three most popular operating systems and their corresponding default filesystems, as well as some challenges in creating code that can work seamlessly across platforms.

# Annex 1

Table 6 shows a summary of the different canonical representations used by some operating systems and the visual representation in the command prompt or the visual explorer tool.

| Name | Input string | NFC | NFD | Mac | Windows explorer | Visual display |
|---|---|---|---|---|---|---|
| Letter A | A (Letter A, 0041) | A | A | A | A | A |
| Letter é | é ( 00E9) | é ( 00E9) | e ' (e + combining accent : 0065, 0301) | é | é | é |
| e followed by combining accent | e ' ( 0065, 0301) | é ( 00E9) | é | é | é | e ' |
| Letter Aring | Å (Aring, 00C5) | Å (A ring, 00C5) | A ° (A + combining ring : 0041, 030A) | Å | Å | Å |
| A + combining ring | A ° (A + combining ring : 0041, 030A) | Å (A ring, 00C5) | A ° (A + combining ring : 0041, 030A) | Å | Å | A ° |
| Letter Aring + cedilla | Å , (Aring + cedilla) | Å , (Aring + cedilla, 00C5, 0327) | A , ° (A + cedilla+ ring : 0041, 0327, 030A) | Å̧ | Å , | Å , |
| Letter A + ring + cedilla | A °, (A + ring + cedilla : 0041, 030A, 0327) | Å , (Aring + cedilla, 00C5, 0327) | A , ° (A + cedilla+ ring : 0041, 0327, 030A) | Å̧ | Å | A °, |
| Letter A + cedilla +ring | A , ° (A + cedilla+ ring : 0041, 0327, 030A) | Å , (Aring + cedilla, 00C5, 0327) | A , ° (A + cedilla+ ring : 0041, 0327, 030A) | Å̧ | Å | A , ° |
| Angstrom | Å (Angstrom, 212B) | Å (A ring, 00C5) | A ° (A + combining ring : 0041, 030A) | Å | Å | ? |
| Angstrom + ring | Å ° (Angstrom, 212B) | | | Å (A with 2 rings stacked vertically) | ? ° | ? ° |
| Angstrom + cedilla | Å , (Angstrom, 212B) | | | Å̧ | ?, | ?, |
| A followed by ligature ffi | A ffi | | | Affi | Affi | A? |

Table 6: Examples of differences in Unicode canonical decomposition

# Annex 2: sample code

### Sample Python code:

```python
import unicodedata
from os import listdir, path

base_dir = '/tmp/py/'

filename = [ u'A', u'A\u00c5', u'A\u030A', u'\u00c5\u0327',
u'A\u030A\u0327', u'A\u0327\u030A', u'\u212b', u'\u212b\u030A',
u'\u212b\u0327', u'a\ufb03', ]
filename_desc = [ 'A\t', 'A-ring', 'A ring', 'A-ring cedil',
'A ring cedil', 'A cedil ring', 'Angstrom','Angstrom ring',
'Angstrom cedl', 'Affi\t' ]

for i in range(len(filename_desc)):
  f2 = base_dir + filename[i]
  with open(f2, "wb") as outfile:
    for j in range(0, (i+1)):
      outfile.write("X")
```

### Sample Java code:

```java
import java.io.File;
import java.io.IOException;
import java.io.PrintWriter;
import java.nio.file.Files;
import java.nio.file.LinkOption;
import java.nio.file.Path;

public class unicode_canonical_paper {

  public static void main(String[] args) {
    String current_folder = "E:\\temp\\";

    String filename1 = "01\u212b.txt";   //Angstrom
    String filename2 = "02\u00c5.txt"; //A-ring character
    String filename3 = "03A\u030A.txt"; //A + ring
    String filename4 = "04c\u00E9.txt";
    String filename5 = "05ce\u0301.txt";

    System.out.println("1 is: "+filename1+ " "+ filename1.length()+" chars   "+
Character.getName((int) filename1.charAt(0)));
    System.out.println("2 is: "+filename2+ " "+ filename2.length()+" chars   "+
Character.getName((int) filename2.charAt(0)));
    System.out.println("3 is: "+filename3+ " "+ filename3.length()+" chars   "+
Character.getName((int) filename3.charAt(0)) + " "+Character.getName((int) filename3.charAt(1)) );
    System.out.println("4 is: "+filename4+ " "+ filename4.length()+" chars   "+
Character.getName((int) filename4.charAt(1))   );
    System.out.println("5 is: "+filename5+ " "+ filename5.length()+" chars   "+
Character.getName((int) filename5.charAt(1))   );

    try{
      PrintWriter pw = new PrintWriter(current_folder+filename1);
      pw.print("1234567890");
      pw.close();
      pw = new PrintWriter(current_folder+filename2);
      pw.print("1234567890");
      pw.close();
      pw = new PrintWriter(current_folder+filename3);
      pw.print("1234567890");
      pw.close();
      pw = new PrintWriter(current_folder+filename4);
      pw.print("1234567890");
      pw.close();
      pw = new PrintWriter(current_folder+filename5);
      pw.print("1234567890");
      pw.close();
    }catch(IOException e){e.printStackTrace(); }
```

```java
    //List files, with java.io
    System.out.println("Listing files in folder " + current_folder + " using java.io.File");
    File folder = new File(current_folder);
            File[] listOfFiles = folder.listFiles();
        for (int i = 0; i < listOfFiles.length; i++) {
          File f = listOfFiles[i];
          read_a_file(f);
        }
  }

  static void read_a_file(File f){
        try {
          System.out.println("\t exists() "+f.exists() + "\n\t isFile() "+ f.isFile() + "\n\t Name:"+ f.getName() + "\\n");// + " " + Character.getName(f.getName().charAt(2)) +

        for(int i=0;i<f.getName().length(); i++)
            System.out.print(Character.getName(f.getName().charAt(i))+ " | ");
        System.out.println("\n\t Path:"+ f.getPath() + " " + Character.getName(f.getPath().charAt(f.getPath().length()-5))
            +"\n\t CanP:"+ f.getCanonicalPath() + " "+ Character.getName(f.getCanonicalPath().charAt(f.getCanonicalPath().length()-5))
            +"\n\t AbsP:"+ f.getAbsolutePath() + " "+ Character.getName(f.getAbsolutePath().charAt(f.getCanonicalPath().length()-5))
            +"\n\t " + f.length());
      } catch (IOException e) {
        e.printStackTrace();
      }
    }
}
```